# Identifying Direct Bandgap Silicon Structures with High-throughput Search and Machine Learning Methods

Rui Wang[1,2†], Hongyu Yu[1,2†], Yang Zhong[1,2*], Hongjun Xiang[1,2*]

[1]Key Laboratory of Computational Physical Sciences (Ministry of Education), Institute of Computational Physical Sciences, State Key Laboratory of Surface Physics, and Department of Physics, Fudan University, Shanghai, 200433, China

[2]Shanghai Qi Zhi Institute, Shanghai, 200030, China

*E-mail: yzhong@fudan.edu.cn; hxiang@fudan.edu.cn

**ABSTRACT:** Utilizations of silicon-based luminescent devices are restricted by the indirect-gap nature of diamond silicon. In this study, the high-throughput method is employed to expedite discoveries of direct-gap silicon crystals. The machine learning (ML) potential is utilized to construct a dataset comprising 2637 silicon allotropes, which is subsequently screened using an ML Hamiltonian model and density functional theory calculations, resulting in identification of 47 direct-gap Si structures. We calculate transition dipole moments (TDM), energies, and phonon bandstructures of these structures to validate their performance. Additionally, we recalculate bandgaps of these structures employing the HSE06 functional. 22 silicon allotropes are identified as potential photovoltaic materials. Among them, the energy per atom of $Si_{22}$-Pm, which has a direct bandgap of 1.27 eV, is 0.026 eV/atom higher than diamond silicon. $Si_{18}$-C2/m, which has a direct bandgap of 0.796 eV, exhibits the highest TDM among identified structures. $Si_{16}$-P$2_1$/c, which has a direct bandgap of 0.907 eV, has the mass density of 2.316 g/cm$^3$, which is the highest among identified structures and higher than that of diamond silicon. The structure $Si_{12}$-P1, which possesses a direct bandgap of 1.69 eV, exhibits the highest spectroscopic limited maximum efficiency (SLME) among identified structures at 32.28%, surpassing that of diamond silicon. This study offers insights into properties of silicon crystals while presenting a systematic high-throughput method for material discovery.

## 1. Introduction

Silicon-based materials have been widely used as photoelectric materials, particularly in the photovoltaic industry[1]. Silicon is the second most abundant element on Earth after oxygen. It is highly stable, non-toxic, and easy to obtain. These characteristics render it a cost-effective material suitable for large-scale production. Moreover, silicon-based PV cells exhibit a conversion efficiency of up to 23%[2] and low energy payback time[3], surpassing most other renewable energy sources.

However, the utilization of silicon-based materials is limited by the indirect band structure of diamond silicon. Consequently, silicon-based luminescent materials have a rather low light emission efficiency, and silicon-based photovoltaic materials need to have sufficient thickness to achieve efficient light absorption[1]. Moreover, due to its direct bandgap of 3.4 eV[4], diamond silicon exhibits a limited capacity for absorbing low-energy photons[5]. Identifying suitable silicon-based photoelectric materials is important. Therefore, significant efforts have been devoted to the search for stable silicon crystals possessing a direct bandgap. Xiang *et al.* employed a particle swarming optimization (PSO) algorithm and predicted a cubic $Si_{20}$ phase with a quasi-direct bandgap of 1.55 eV[6]. Wang *et al.* predicted six metastable silicon crystals which exhibit direct or quasi-direct bandgaps ranging from 0.39-1.25 eV[7]. Lee *et al.* utilized a combination of the conformational space annealing method and the global optimization method to predict eight direct bandgap silicon semiconductor materials[8]. Guo *et al.* proposed a new silicon phase that possesses a direct band gap of 0.61 eV and exhibits dynamic, thermal, and mechanical stability, demonstrating great potential as a photoelectric material[9]. Zhang *et al.* found four silicon allotropes with direct or quasi-direct ranging from 1.193 to 1.473 eV[5]. He *et al.* employed *ab initio* random structure search[10,11] and random sampling strategy combined with space group and graph theory[12] and discovered two direct bandgap silicon structures, Pbam-32 and P6/mmm, with direct bandgaps of 1.39 eV and 1.13 eV, respectively[13]. Despite the discovery of these direct bandgap Si structures, high-performance silicon-based lasers have not yet been realized experimentally. However, there may still exist other metastable Si structures with superior photoelectric properties, greater stability, and better suitability for photovoltaic applications yet to be uncovered.

High-throughput computation has emerged as a powerful tool for accelerating the discovery and design of novel materials with enhanced properties through the rapid screening and prediction of the properties of thousands of materials[14–16]. The high-throughput approach involves the creation of large databases of materials, often containing millions of potential candidates. The use of massive computational resources and machine-learning algorithms enables scientists to explore a vast materials design space, generating a wealth of data that can be used to guide experimental work and accelerate the discovery of new materials. This methodology has been applied to a diverse range of materials, from catalysts[17] and batteries[18] to semiconductors[19] and thermoelectrics[20].

In this study, we propose an efficient and rapid approach for identifying direct bandgap silicon crystals by utilizing a high-throughput computation method combined with machine learning potential and Hamiltonian. We created a dataset of Si crystals by substituting the atoms in the structures from the Carbon-24 dataset[21] with silicon atoms and proportionally scaling the lattice. Subsequently, we optimized the structure using the neuroevolution machine-learned potential (NEP) of GPUMD[22]. The dataset comprises 2637 distinct Si crystal structures that have been optimized with machine learning potentials. Using the HamGNN model[23], a transferable E(3) equivariant graph neural network (GNN) that can predict the electronic Hamiltonian matrices of molecules and solids, we calculated the band structures of all the silicon crystals in the dataset and identified 47 direct bandgap structures. Among them, 24 candidates are both thermodynamically and dynamically stable with non-zero transition dipole moments at the direct gap. To obtain more accurate values of the bandgaps, we recalculated them using the HSE06 functional, and 22 of these structures exhibit direct bandgaps. By identifying these Si crystals, our study provides a valuable reference for researchers interested in developing high-performance Si-based photoelectric devices. Also, the workflow we employed is not only applicable to silicon-based materials but can also extend to other materials.

## 2. Methods

### 2.1. Details for the optimization process utilizing machine learning potentials

This study utilized the machine learning potential GPUMD to optimize the initial Si structures. The GPUMD potential was trained on a Si dataset comprising 2475 Si structures[24], which included various Si crystal structures, diamond surfaces, diamond structures with vacancies, amorphous Si structures, and some other Si structures. The convergence criterion during the optimization process was that the forces on each atom be less than 0.01 eV/Å, ensuring the resulting structures were at a local energy minimum.

### 2.2. Details for the training of the HamGNN model

This study employed the machine learning Hamiltonian model HamGNN[23] to predict the band structures of silicon allotropes. The model was trained on 30 silicon allotropes from the Materials Project[25]. The Hamiltonian matrices of these silicon allotropes were computed through DFT calculations using OpenMX[26]. The dataset was divided into training, validation, and test sets with a ratio of 0.8:0.1:0.1. The HamGNN model was first trained with solely the loss value of the Hamiltonian until it reached convergence. Then the model continued to be trained with both the loss value of the Hamiltonian and the band energy in the loss function. The MAE of the Hamiltonian matrix predicted by HamGNN for silicon allotropes in the test set is only 2.01 meV, demonstrating that HamGNN achieves high precision even with relatively small datasets. After the two training steps, the final model is used for predicting the electronic structure of various silicon candidates.

### 2.3. Details of the DFT calculations

We conducted DFT calculations on the structural optimizations, transition dipole moments, phonon band structures, and energies on the silicon structures utilizing the Vienna ab initio Simulation package (VASP) with the post-processing VASPKIT package[27]. For the exchange-correlation potential, we employed the generalized gradient approximation (GGA) within the Perdew-Burke-Ernzerhof (PBE) functional[28]. The energy convergence criterion was set to $10^{-8}$ eV. The plane wave cutoff was 400 eV. The Brillouin zone integrations were carried out using a dense K-point mesh within the Γ scheme with a spacing of $2\pi \times 0.02$ Å$^{-1}$ in the reciprocal cell. We employed the supercell method implemented in the Phonopy package[29] to calculate the phonon band structures of the silicon structures. When performing band structure calculations for silicon structures, we first used the PBE functional with an energy convergence criterion of $10^{-8}$ eV and a plane wave cutoff of 400 eV to screen the direct band gap structures. Then, we employed the HSE06 functional[30] with an energy convergence criterion of $10^{-8}$ eV and a plane wave cutoff of 500 eV to obtain a more accurate value for the bandgap.

## 3. Results and discussions

### 3.1. High-throughput search steps

High-throughput material design typically involves creating a dataset and screening materials. Dataset construction involves compiling relevant material data. Material screening involves the utilization of computational models to identify materials possessing desired properties. In this study, we have constructed a dataset for Si from the Carbon-24 dataset, which comprises potential direct bandgap Si structures. Then, we utilized HamGNN, an ML Hamiltonian model, to identify direct bandgap Si structures in the dataset. Finally, we validated the results via DFT-based calculations.

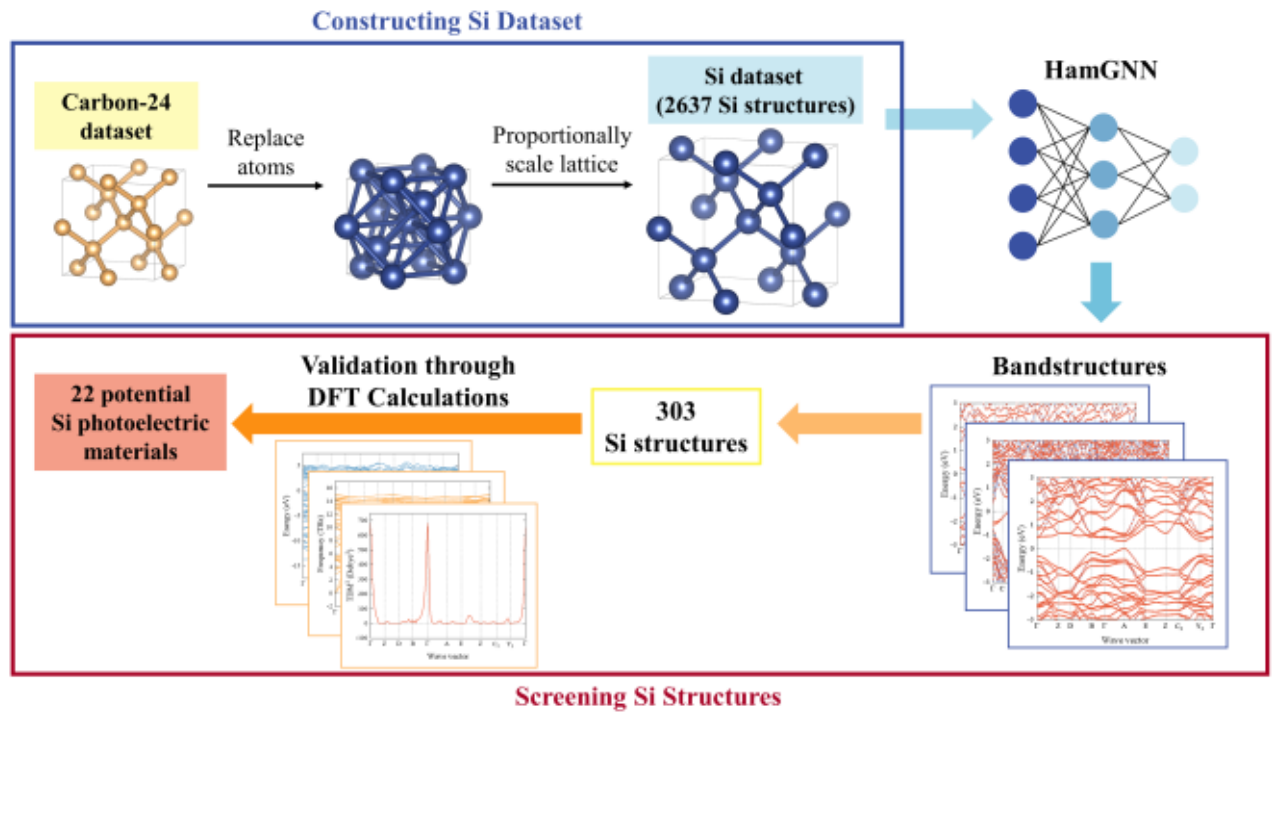

Figure 1: Pipeline of finding potential photoelectric Si-based materials through high-throughput computation method.

### 3.1.1. Construction of Si Dataset

The Si dataset used in this paper derives from the Carbon-24 dataset, given that Si and C atoms are both tetravalent. Carbon-24 dataset contains 10,153 carbon allotropes, each with 6 to 24 atoms in its unit cell. These structures are at a local energy minimum after DFT-based relaxation. The Si dataset utilized in this article was generated by substituting C atoms with Si atoms in the crystal structures of the Carbon-24 dataset to produce initial Si structures[31,32]. Next, to achieve reasonable lattices, proportional scaling of the lattice was carried out. Considering different bond lengths between C-C atoms (158 pm) and Si-Si atoms (230 pm), the initial lattice structures were enlarged by a factor of 1.46, as depicted in Figure 1. Finally, machine learning potentials were used to optimize the initial Si structures. Specifically, the neuroevolution machine-learned potential (NEP) of GPUMD[22,33] was employed, and the optimization was accomplished using the PyNEP package[33]. The RMSE of energies and RMSE of forces of the GPUMD potential are 5.7 meV/atom and 107 meV/Å[22], respectively. Similar structures were removed from the resulting dataset, resulting in a Si dataset consisting of 2637 distinct and stable Si structures. By substituting C atoms in structures from the Carbon-24 dataset with Si atoms, the resulting dataset is more comprehensive, which serves as an expansion of the existing Si dataset.

### 3.1.2. Screening Structures with Direct Bandgap using HamGNN

To identify Si structures with direct bandgap, we used the HamGNN model[23] for initial screening. HamGNN is an E(3) equivariant deep neural network that can predict the electronic Hamiltonian matrices of solids. By employing E(3) equivariant deep neural networks, HamGNN obeys all the fundamental equivariant constraints of Hamiltonians and achieves high transferability across diverse datasets. We utilized a HamGNN model that was trained on 30 silicon allotropes obtained from the Materials Project[25]. We tested the accuracy of the resulting model on 9 additional silicon allotropes from the Materials Project[25]. These structures contain between 68 and 232 atoms in their unit cells. The MAE of the Hamiltonian matrix predicted by HamGNN for the silicon structures in the test set is 2.43meV. As illustrated in Figure 2, the HamGNN model exhibits high accuracy in comparison to results obtained from OpenMX calculations[26].

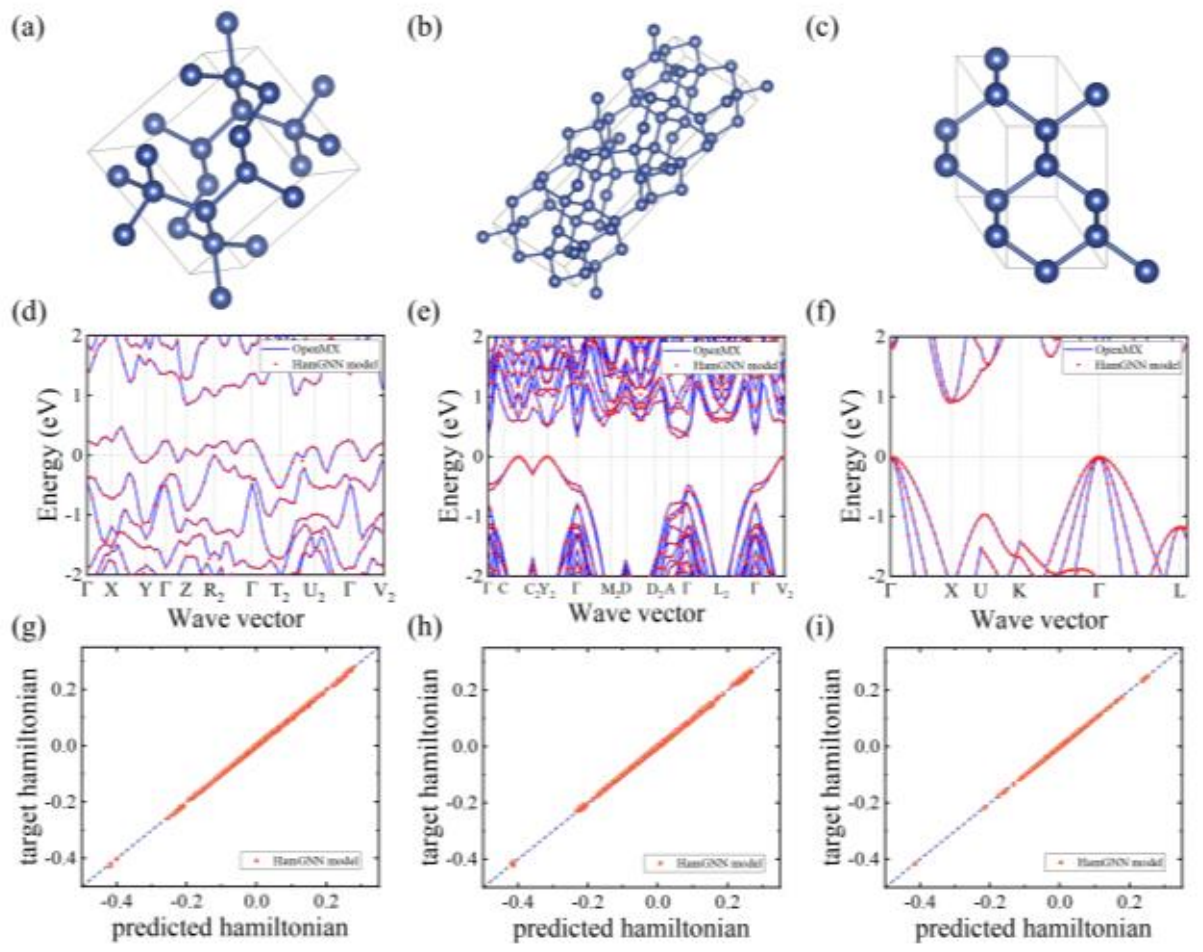

Figure 2: Comparison of the band structures and Hamiltonian matrix of (a) $Si_{10}$, (b) $Si_{24}$, and (c) $Si_6$, predicted by the HamGNN model (dotted lines) and OpenMX (solid lines). The structures were selected from the generated Si dataset.

The high efficiency of HamGNN enables us to efficiently acquire the band structures of Si structures. We employed the HamGNN model to determine the direct bandgap and indirect bandgap of the structures. The calculation of the Hamiltonian matrix for a single structure required only 0.22 seconds on average, which is substantially faster compared to conventional methods. The direct bandgap is the energy difference between the top of the valence band and the bottom of the conduction band at the same k-point. The indirect bandgap is computed between the maximum of the valence band and the minimum of the conduction band at different k points in reciprocal space. The HamGNN calculations enable rapid identification of crystal structures exhibiting direct bandgaps. A difference of less than 0.1 eV between the direct and indirect bandgaps serves as a criterion for identifying crystal structures that exhibit a direct bandgap. Employing the HamGNN model, a total of 303 crystal structures from the Si dataset are estimated to have a direct bandgap.

The HamGNN model we employed was trained on Density Functional Theory (DFT) calculations utilizing the Perdew-Burke-Ernzerhof (PBE) functional[28]. Consequently, when validating these structures through DFT calculations, we initially employed the PBE functional to screen the structures and subsequently utilized the HSE06 functional[30] to acquire a more precise value of the bandgap.

### 3.1.3. Validation through Density Functional Theory

The Si crystals produced in the previous step mostly exhibit P1 symmetry. To ensure the accuracy of our calculations, we initially symmetrized the structures in the Si dataset. This procedure helped to eliminate potential errors caused by the lack of symmetry in the crystal structures. We then performed structure relaxation using density functional theory (DFT) calculations to find the most stable configurations of these structures. Finally, we calculated the band structures of the crystals obtained from the previous step. The electronic band calculations were performed with density functional theory (DFT) by combining the Vienna ab initio Simulation package (VASP) with the post-processing VASPKIT package[27]. For the exchange-correlation potential, we considered the generalized gradient approximation (GGA) within the Perdew-Burke-Ernzerhof (PBE) functional[28]. The energy convergence criterion was set to $10^{-8}$ eV. The Brillouin zone integrations were carried out using a dense K-point mesh within the Γ scheme. As a result, we found 47 crystal structures with a direct bandgap, indicating their potential suitability for optoelectronic applications. Additionally, 78 structures exhibit quasi-direct bandgaps, with a difference of less than 0.1 eV between the direct and indirect bandgaps[8,34,35]. Scatter plot of the energies against volumes of the direct, quasi-direct and indirect bandgap structures can be found in the supporting information.

To further investigate the potential of the 47 Si structures as photovoltaic materials, we used the VASPKIT package[27] to calculate the transition dipole moment (TDM) between the conduction band and the valence band. As a result, we identified 27 Si structures where the transition is allowed from the valence band to the conduction band. Additionally, we employed the VASP code and the Phonopy package[29] to calculate the energy and the phonon band structures of the Si structures. We identified 24 direct bandgap Si structures that are dynamically stable.

Typically, the band gaps obtained through DFT calculations are underestimated by 30-50%, indicating that the true band gap value is higher than the calculated results. Heyd *et al.* proposed the HeydScuseria-Ernzerhof (HSE06) functional, which is a more feasible hybrid functional method for addressing this issue[30]. To obtain a more accurate bandgap value, we recalculated the bandgap of the 24 structures employing HSE06 functional and found that 22 of them exhibit direct bandgaps. A combination of machine learning potential and Hamiltonian and DFT calculations leads to much less computation resource cost and high efficiency in finding ideal silicon with direct bandgaps and stable structure.

## 3.2. High-throughput search results

### 3.2.1. The Si dataset

In this study, we generated a comprehensive dataset of silicon-based crystal structures by using machine learning potential to optimize initial structures obtained from the Carbon-24 dataset. We obtained a Si dataset of 2637 distinct crystal structures. The optimization process ensured that the structures were stable and reasonable, providing a reliable dataset for further study. This new Si dataset is a significant extension of the existing dataset of silicon-based crystal structures. To validate the uniqueness of our dataset, we compared it with an existing dataset[24], which consists of 2475 Si structures. We found that only 7 structures were alike in the two datasets. This result emphasizes the importance of our newly generated dataset, which is a significant expansion to the currently known silicon-based crystal structures. Therefore, this new dataset has the potential to stimulate future research and support the design of new silicon-based materials for technological applications.

Our workflow for generating a novel dataset, which includes creating initial structures from existing datasets, proportionally scaling the crystal lattice, and optimizing these structures through machine learning potentials, is not exclusive to Si. We suggest that it could be appropriate for creating new datasets of various other materials. As an instance, this same approach could be utilized for producing a new dataset of Ge. This workflow of generating new datasets offers an efficient and inexpensive approach to investigating new materials.

### 3.2.2. New Si structures with direct gaps

Using DFT calculation, we identified 47 Si crystals with direct bandgaps. To further investigate the potential of the 47 Si structures as photovoltaic materials, we used the VASPKIT package[27] to calculate the transition dipole moment (TDM) between the highest conduction band and the lowest valence band. We observed that among the 47 crystal structures examined, a total of 27 exhibited non-zero transition dipole moments at the direct gap, indicating an allowed transition from the valence band to the conduction band and demonstrating its suitability for light-based electronic devices. To further validate the applicability of the identified Si crystals, we conducted additional calculations to determine the energy of the structures and analyze their phonon band structures. We applied the same calculations to all 27 Si structures exhibiting nonzero TDMs at the direct gap and discovered that among them, a total of 24 structures are dynamically stable. The diamond Si is widely recognized as the most thermodynamically stable silicon structure. Accordingly, we conducted a comparative analysis of the energies between the identified Si structures and diamond Si. The obtained results provide crucial insights into the stability of these Si crystals. The bandgap of the 24 structures was recalculated using the HSE06 functional to obtain a more precise value. We found that 22 of them exhibit direct bandgap, and their corresponding energies and bandgaps are listed in Table 1. POSCAR files of these structures can be found in the supporting information. The results demonstrate the potential of Si crystals for light-based electronic applications. These findings provide a foundation for further research aiming to design and develop novel Si-based optoelectronic devices that offer improved efficiency and performance.

In the 22 identified structures, the coordination number of silicon atoms is consistently 4. In these structures, bond angles vary from 59.64° to 142.13°, and Si-Si bond lengths range between 2.27 Å andna 2.52 Å. Additionally, to demonstrate the effectiveness of our method, we compared the Si structures resulting from our study with the direct bandgap silicon crystals reported in previous research. We found that four of the Si structures we discovered were already reported in other papers[8,36–38], substantiating the reliability of our methodology. Furthermore, the overall energy of our structures is lower than the previous silicon crystals found, suggesting that the Si structures we discovered are relatively stable. Our findings suggest that these 22 Si crystals are promising candidates for further experimental investigation and potential utilization in optoelectronic devices. These metastable Si structures could be grown using techniques such as epitaxial growth or ion implantation[34].

Figure 3(a) illustrates one of the identified structures, $Si_{18}$, which possesses C2/m symmetry and exhibits a bandgap energy of 0.796eV. The bond angles in this structure range between 91.90° to 127.37°, and Si-Si bond lengths vary from 2.30 Å and 2.52 Å. Its $TDM^2$ at the bandgap is 7442.967 $Debye^2$. Its energy is 0.056 eV/atom higher than diamond Si. Figure 3(d) depicts the absorption coefficient of $Si_{18}$. $Si_{18}$ has the largest $TDM^2$ at the bandgap among all the identified structures, resulting in a significantly enhanced absorption coefficient compared to diamond Si[39]. This renders it well-suited for photovoltaic applications.

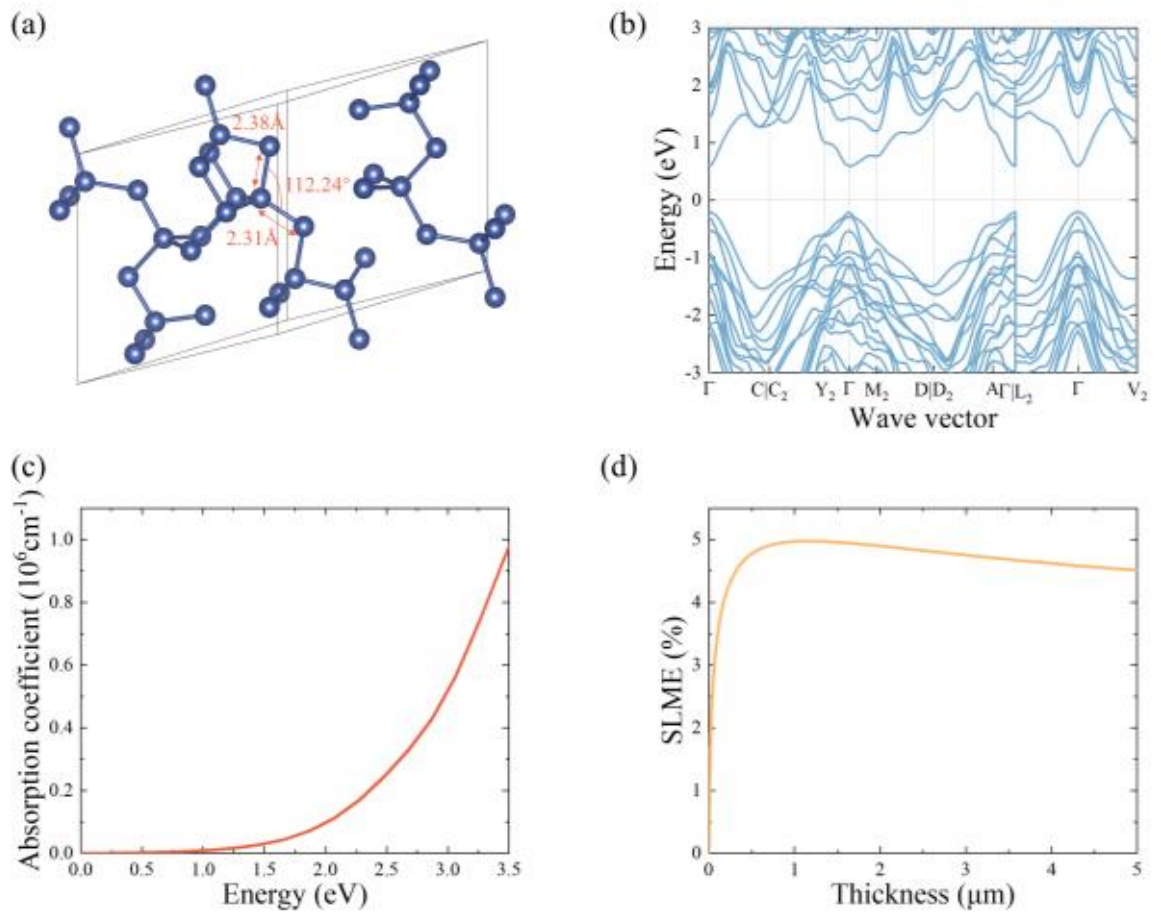

Figure 3. The (a) unit cell, (b) band structure, (c) phonon band structure and (d) absorption coefficient of $Si_{18}$.

Figure 4(a) illustrates one of the identified structures, $Si_{16}$. It belongs to $P2_1/c$ symmetry group and exhibits a direct bandgap of 0.907eV. Si-Si bond lengths in this structure vary from 2.31 Å to 2.46 Å, and the bond angles range from 95.46° to 128.91°. Its energy is 0.111eV/atom higher than that of diamond Si. $Si_{16}$ has the highest mass density, 2.316g/$cm^3$, among all identified structures. It is higher than that of diamond Si, which is 2.313g/$cm^3$.

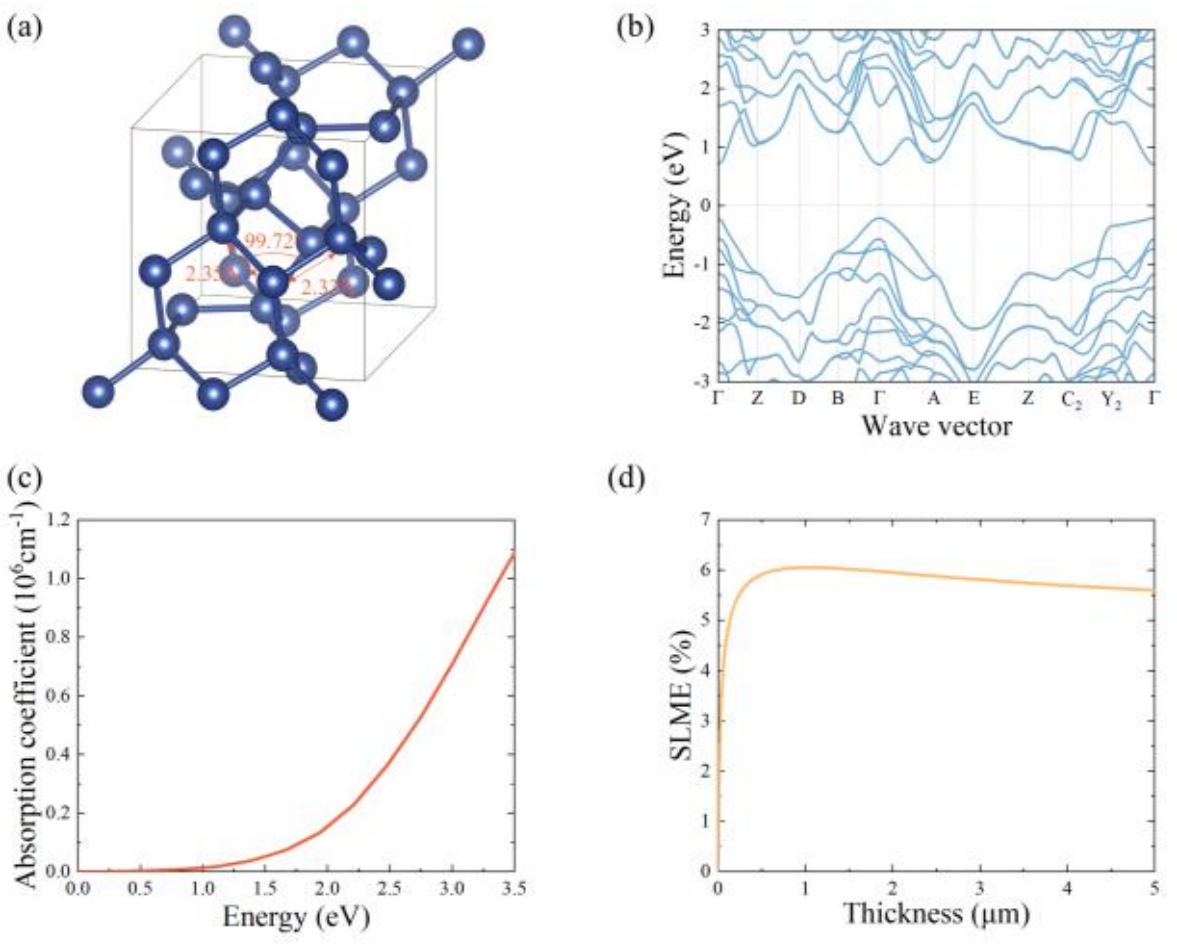

Figure 4. The (a) unit cell, (b) band structure, (c) phonon band structure and (d) absorption coefficient of $Si_{16}$.

Figure 5 illustrates the unit cell, band structure, absorption coefficient, and SLME of $Si_{22}$. $Si_{22}$ is a superlattice of diamond Si and possesses the lowest energy among all the identified structures. Remarkably, its energy per atom is 0.026 eV/atom higher than that of diamond Si, which is lower than the energies reported for most other direct bandgap Si structures found in previous research[5–9,36–38,40]. This energy is slightly higher than that of M585-silicon, which is discovered by He *et al.* and processes an energy per atom 0.025 eV/atom higher than that of diamond Si[31]. We calculated the spectroscopic limited maximum efficiency (SLME)[41] of $Si_{22}$ utilizing the PBE functional. The SLME of the $Si_{22}$ structure is 22.23%. The $Si_{22}$ structure exhibits Pm symmetry and a bandgap energy of 1.27eV. In this structure, bond angles vary from 90.20° to 135.17°, and Si-Si bond lengths range between 2.33 Å and 2.40 Å.

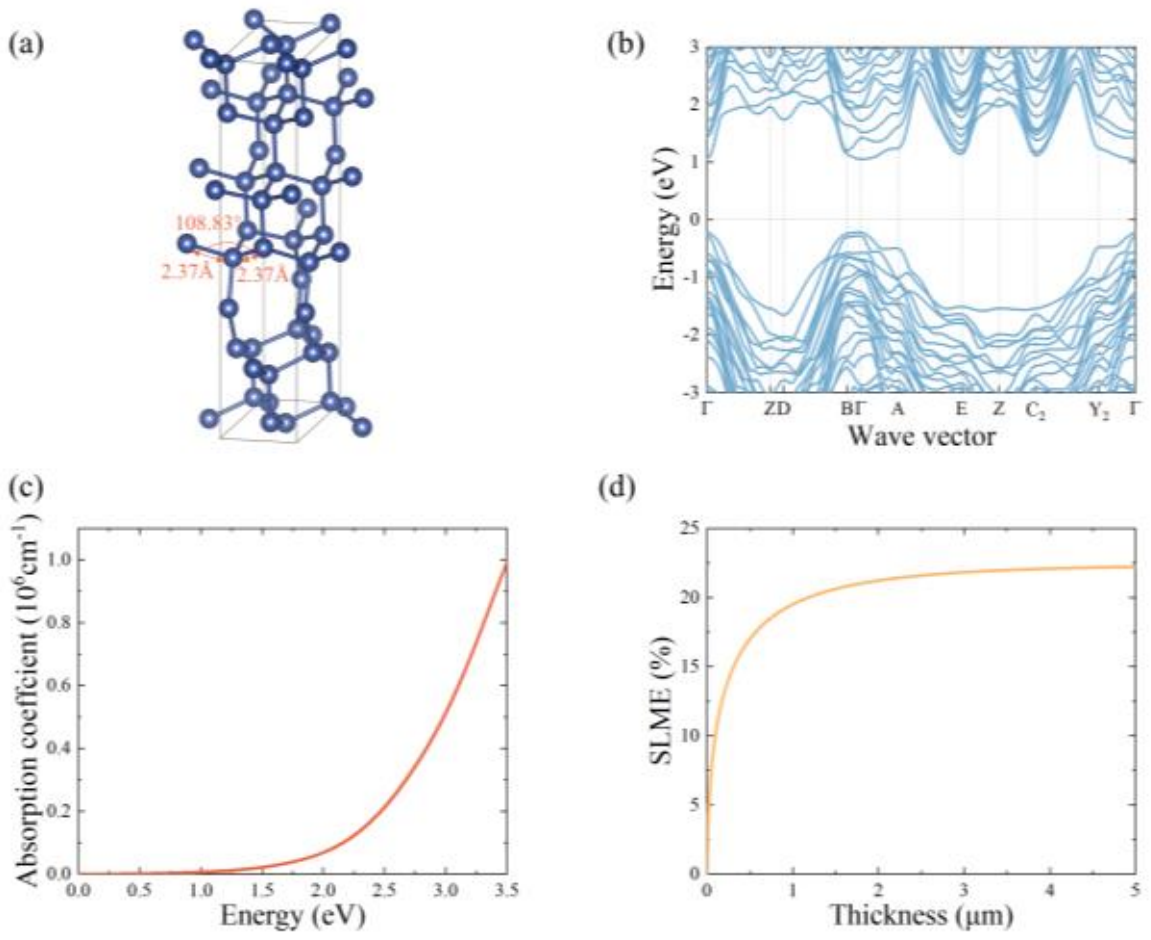

Figure 5. The (a) unit cell, (b) band structure, (c) absorption coefficient and (d) SLME of $Si_{22}$.

Figure 6 depicts the identified structure, $Si_{12}$, which has the highest SLME among the identified structures. The SLME of $Si_{12}$ is 32.28%, which is higher than that of diamond silicon and previously reported direct bandgap silicon structures[8]. The high SLME of $Si_{12}$ renders it a suitable candidate for photovoltaic applications. The $Si_{12}$ structure exhibits P1 symmetry, and its energy per atom is 0.116 eV/atom above that of diamond silicon. It exhibits a direct bandgap of 1.69 eV. The bond angles in this structure range from 86.95° to 136.11°, while the Si-Si bond lengths vary between 2.33 Å and 2.44 Å.

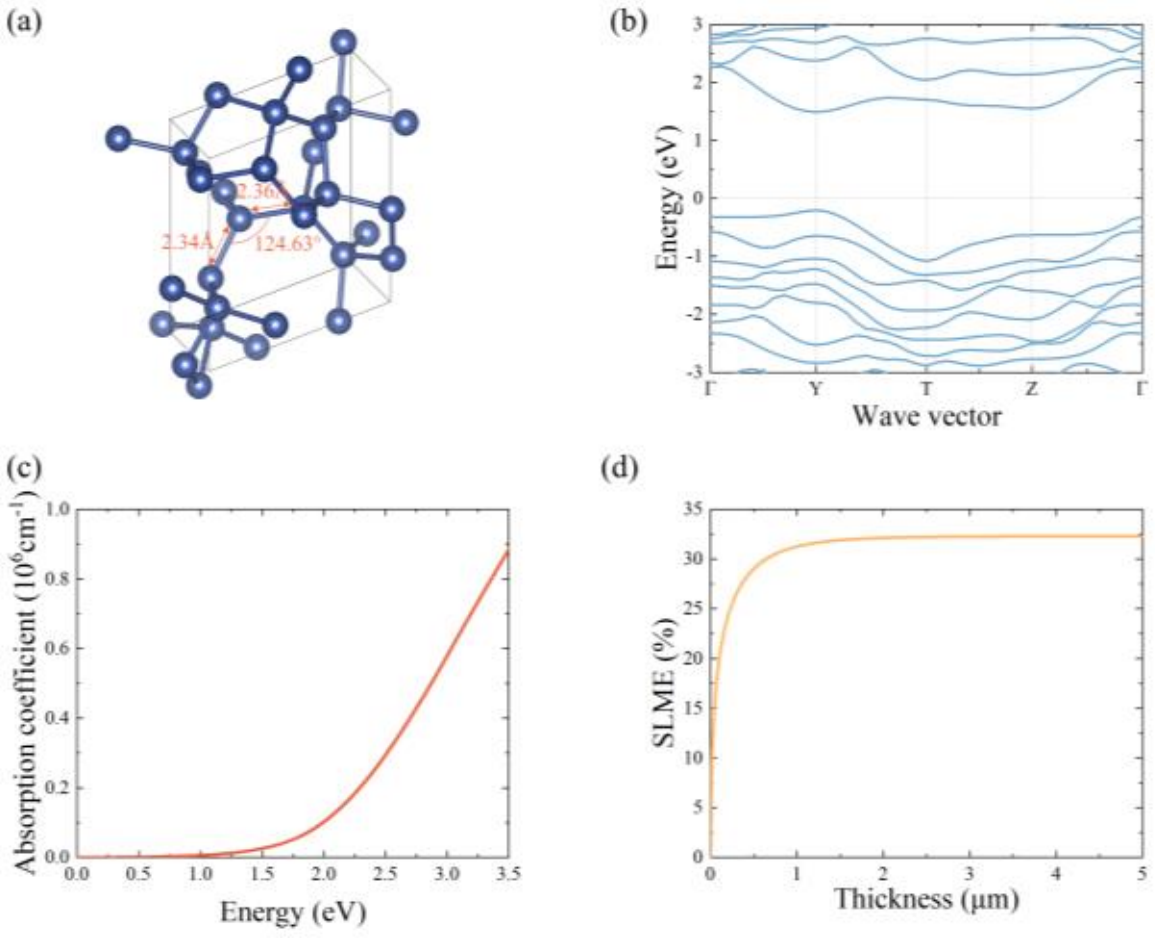

Figure 6: The (a) unit cell, (b) band structure, (c) absorption coefficient and (d) SLME of $Si_{12}$.

## 4. Conclusion

In summary, this study produced a new dataset of silicon crystals and employed high-throughput computation to identify 22 potential photovoltaic materials. The initial structures of the dataset were created from pre-existing datasets and were then subjected to proportional scaling of the crystal lattice. The structures were then optimized using machine learning potentials, resulting in 2637 distinct and stable Si structures. We employed the HamGNN machine learning Hamiltonian model and DFT calculations to screen the dataset. Furthermore, we performed additional calculations to obtain the transition dipole moment, energy, and phonon band structures of these structures. As a result, we identified 22 structures that have the potential to serve as efficient photovoltaic materials. Among these structures, the energy per atom of $Si_{22}$ is only 0.026 eV/atom higher than that of diamond silicon. $Si_{18}$ exhibits the highest $TDM^2$ at the direct bandgap of the identified structures, which is 7442.967 $Debye^2$. $Si_{16}$ has the largest mass density among the identified structures, which is 2.316g/$cm^3$, higher than that of diamond silicon. The SLME of $Si_{12}$ is found to be the highest among the identified structures, reaching 32.28%, surpassing previously reported direct bandgap silicon structures. Overall, our findings offer significant insights into the photoelectric properties of silicon crystals and their potential in optoelectronic applications. Furthermore, the workflow we employed is not limited to silicon materials but can be extended to other types of materials as well. It provides a valuable tool for exploring a broad range of materials and their corresponding properties.

## ASSOCIATED CONTENT

**Supporting Information**.

- Scatterplot of the energies (eV/atom) calculated using PBE functionals against volumes (Å3) of the structures identified utilizing HamGNN.

This material is available free of charge via the Internet at http://pubs.acs.org.

## AUTHOR INFORMATION

### Corresponding Author

Yang Zhong - Key Laboratory of Computational Physical Sciences (Ministry of Education), Institute of Computational Physical Sciences, State Key Laboratory of Surface Physics, and Department of Physics, Fudan University, Shanghai, 200433, China; Shanghai Qi Zhi Institute, Shanghai, 200030, China; Email: yzhong@fudan.edu.cn.

Hongjun Xiang - Key Laboratory of Computational Physical Sciences (Ministry of Education), Institute of Computational Physical Sciences, State Key Laboratory of Surface Physics, and Department of Physics, Fudan University, Shanghai, 200433, China; Shanghai Qi Zhi Institute, Shanghai, 200030, China; Email: hxiang@fudan.edu.cn.

### Authors

Rui Wang - Key Laboratory of Computational Physical Sciences (Ministry of Education), Institute of Computational Physical Sciences, State Key Laboratory of Surface Physics, and Department of Physics, Fudan University, Shanghai, 200433, China; Shanghai Qi Zhi Institute, Shanghai, 200030, China.

Hongyu Yu - Key Laboratory of Computational Physical Sciences (Ministry of Education), Institute of Computational Physical Sciences, State Key Laboratory of Surface Physics, and Department of Physics, Fudan University, Shanghai, 200433, China; Shanghai Qi Zhi Institute, Shanghai, 200030, China.

### Author Contributions

The manuscript was written through contributions of all authors. Rui Wang* and Hongyu Yu* contributed equally to this work. All authors have given approval to the final version of the manuscript.

## ACKNOWLEDGMENT

We acknowledge financial support from the National Key R&D Program of China (No. 2022YFA1402901), NSFC (grants No. 11991061, 12188101), Shanghai Science and Technology Program (No. 23JC1400900), and the Guangdong Major Project of the Basic and Applied Basic Research (Future functional materials under extreme conditions--2021B0301030005).

## REFERENCES

(1) Ballif, C.; Haug, F.-J.; Boccard, M.; Verlinden, P. J.; Hahn, G. Status and Perspectives of Crystalline Silicon Photovoltaics in Research and Industry. *Nat. Rev. Mater.* **2022**, *7* (8), 597–616. https://doi.org/10.1038/s41578-022-00423-2.

(2) Bullock, J.; Wan, Y.; Hettick, M.; Zhaoran, X.; Phang, S. P.; Yan, D.; Wang, H.; Ji, W.; Samundsett, C.; Hameiri, Z.et al. Dopant-Free Partial Rear Contacts Enabling 23% Silicon Solar Cells. *Adv. Energy Mater.* **2019**, *9* (9), 1803367. https://doi.org/10.1002/aenm.201803367.

(3) Bhandari, K. P.; Collier, J. M.; Ellingson, R. J.; Apul, D. S. Energy Payback Time (EPBT) and Energy Return on Energy Invested (EROI) of Solar Photovoltaic Systems: A Systematic Review and Meta-Analysis. *Renewable Sustainable Energy Rev.* **2015**, *47*, 133–141. https://doi.org/10.1016/j.rser.2015.02.057.

(4) Hybertsen, M. S.; Louie, S. G. First-Principles Theory of Quasiparticles: Calculation of Band Gaps in Semiconductors and Insulators. *Phys. Rev. Lett.* **1985**, *55* (13), 1418–1421. https://doi.org/10.1103/PhysRevLett.55.1418.

(5) Zhang, W.; Chai, C.; Fan, Q.; Song, Y.; Yang, Y. Direct and Quasi-Direct Band Gap Silicon Allotropes with Low Energy and Strong Absorption in the Visible for Photovoltaic Applications. *Results Phys.* **2020**, *18*, 103271. https://doi.org/10.1016/j.rinp.2020.103271.

(6) Xiang, H. J.; Huang, B.; Kan, E.; Wei, S.-H.; Gong, X. G. Towards Direct-Gap Silicon Phases by the Inverse Band Structure Design Approach. *Phys. Rev. Lett.* **2013**, *110* (11), 118702. https://doi.org/10.1103/PhysRevLett.110.118702.

(7) Wang, Q.; Xu, B.; Sun, J.; Liu, H.; Zhao, Z.; Yu, D.; Fan, C.; He, J. Direct Band Gap Silicon Allotropes. *J. Am. Chem. Soc.* **2014**, *136* (28), 9826–9829. https://doi.org/10.1021/ja5035792.

(8) Lee, I.-H.; Lee, J.; Oh, Y. J.; Kim, S.; Chang, K. J. Computational Search for Direct Band Gap Silicon Crystals. *Phys. Rev. B* **2014**, *90* (11), 115209. https://doi.org/10.1103/PhysRevB.90.115209.

(9) Guo, Y.; Wang, Q.; Kawazoe, Y.; Jena, P. A New Silicon Phase with Direct Band Gap and Novel Optoelectronic Properties. *Sci. Rep.* **2015**, *5* (1), 14342. https://doi.org/10.1038/srep14342.

(10) Pickard, C. J.; Needs, R. J. High-Pressure Phases of Silane. *Phys. Rev. Lett.* **2006**, *97* (4), 045504. https://doi.org/10.1103/PhysRevLett.97.045504.
(11) Pickard, C. J.; Needs, R. J. Ab Initio Random Structure Searching. *J. Phys.: Condens. Matter* **2011**, *23* (5), 053201. https://doi.org/10.1088/0953-8984/23/5/053201.
(12) Shi, X.; He, C.; Pickard, C. J.; Tang, C.; Zhong, J. Stochastic Generation of Complex Crystal Structures Combining Group and Graph Theory with Application to Carbon. *Phys. Rev. B* **2018**, *97* (1), 014104. https://doi.org/10.1103/PhysRevB.97.014104.
(13) He, C.; Shi, X.; Clark, S. J.; Li, J.; Pickard, C. J.; Ouyang, T.; Zhang, C.; Tang, C.; Zhong, J. Complex Low Energy Tetrahedral Polymorphs of Group IV Elements from First Principles. *Phys. Rev. Lett.* **2018**, *121* (17), 175701. https://doi.org/10.1103/PhysRevLett.121.175701.
(14) Curtarolo, S.; Hart, G. L. W.; Nardelli, M. B.; Mingo, N.; Sanvito, S.; Levy, O. The High-Throughput Highway to Computational Materials Design. *Nat. Mater.* **2013**, *12* (3), 191–201. https://doi.org/10.1038/nmat3568.
(15) Gubaev, K.; Podryabinkin, E. V.; Hart, G. L. W.; Shapeev, A. V. Accelerating High-Throughput Searches for New Alloys with Active Learning of Interatomic Potentials. *Computational Materials Science* **2019**, *156*, 148–156. https://doi.org/10.1016/j.commatsci.2018.09.031.
(16) Choudhary, K.; Garrity, K. F.; Sharma, V.; Biacchi, A. J.; Hight Walker, A. R.; Tavazza, F. High-Throughput Density Functional Perturbation Theory and Machine Learning Predictions of Infrared, Piezoelectric, and Dielectric Responses. *npj Comput Mater* **2020**, *6* (1), 1–13. https://doi.org/10.1038/s41524-020-0337-2.
(17) Sehested, J.; Larsen, K. E.; Kustov, A. L.; Frey, A. M.; Johannessen, T.; Bligaard, T.; Andersson, M. P.; Nørskov, J. K.; Christensen, C. H. Discovery of Technical Methanation Catalysts Based on Computational Screening. *Top. Catal.* **2007**, *45* (1), 9–13. https://doi.org/10.1007/s11244-007-0232-9.
(18) Kabiraj, A.; Mahapatra, S. High-Throughput First-Principles-Calculations Based Estimation of Lithium Ion Storage in Monolayer Rhenium Disulfide. *Commun. Chem.* **2018**, *1* (1), 1–10. https://doi.org/10.1038/s42004-018-0082-3.
(19) Guha, S.; Kabiraj, A.; Mahapatra, S. High-Throughput Design of Functional-Engineered MXene Transistors with Low-Resistive Contacts. *npj Comput. Mater.* **2022**, *8* (1), 1–13. https://doi.org/10.1038/s41524-022-00885-6.
(20) Gan, Y.; Wang, G.; Zhou, J.; Sun, Z. Prediction of Thermoelectric Performance for Layered IV-V-VI Semiconductors by High-Throughput Ab Initio Calculations and Machine Learning. *npj Comput. Mater.* **2021**, *7* (1), 1–10. https://doi.org/10.1038/s41524-021-00645-y.
(21) Pickard; J., C. AIRSS Data for Carbon at 10GPa and the C+N+H+O System at 1GPa, 2020. https://doi.org/10.24435/MATERIALSCLOUD:2020.0026/V1.
(22) Fan, Z.; Zeng, Z.; Zhang, C.; Wang, Y.; Song, K.; Dong, H.; Chen, Y.; Ala-Nissila, T. Neuroevolution Machine Learning Potentials: Combining High Accuracy and Low Cost in Atomistic Simulations and Application to Heat Transport. *Phys. Rev. B* **2021**, *104* (10), 104309. https://doi.org/10.1103/PhysRevB.104.104309.
(23) Zhong, Y.; Yu, H.; Su, M.; Gong, X.; Xiang, H. Transferable Equivariant Graph Neural Networks for the Hamiltonians of Molecules and Solids. *npj Comput Mater* **2023**, *9* (1), 182. https://doi.org/10.1038/s41524-023-01130-4.
(24) Bartók, A. P.; Kermode, J.; Bernstein, N.; Csányi, G. Machine Learning a General-Purpose Interatomic Potential for Silicon. *Phys. Rev. X* **2018**, *8* (4), 041048. https://doi.org/10.1103/PhysRevX.8.041048.
(25) Jain, A.; Ong, S. P.; Hautier, G.; Chen, W.; Richards, W. D.; Dacek, S.; Cholia, S.; Gunter, D.; Skinner, D.; Ceder, G. et al. Commentary: The Materials Project: A Materials Genome Approach to Accelerating Materials Innovation. *APL Mater.* **2013**, *1* (1), 011002. https://doi.org/10.1063/1.4812323.
(26) Ozaki, T. Variationally Optimized Atomic Orbitals for Large-Scale Electronic Structures. *Phys. Rev. B* **2003**, *67* (15), 155108. https://doi.org/10.1103/PhysRevB.67.155108.
(27) Wang, V.; Xu, N.; Liu, J.-C.; Tang, G.; Geng, W.-T. VASPKIT: A User-Friendly Interface Facilitating High-Throughput Computing and Analysis Using VASP Code. *Comput. Phys. Commun.* **2021**, *267*, 108033. https://doi.org/10.1016/j.cpc.2021.108033.
(28) Perdew, J. P.; Burke, K.; Ernzerhof, M. Generalized Gradient Approximation Made Simple [Phys. Rev. Lett. 77, 3865 (1996)]. *Phys. Rev. Lett.* **1997**, *78* (7), 1396–1396. https://doi.org/10.1103/PhysRevLett.78.1396.
(29) Togo, A. First-Principles Phonon Calculations with Phonopy and Phono3py. *J. Phys. Soc. Jpn.* **2023**, *92* (1), 012001. https://doi.org/10.7566/JPSJ.92.012001.
(30) Krukau, A. V.; Vydrov, O. A.; Izmaylov, A. F.; Scuseria, G. E. Influence of the Exchange Screening Parameter on the Performance of Screened Hybrid Functionals. *J. Chem. Phys.* **2006**, *125* (22), 224106. https://doi.org/10.1063/1.2404663.
(31) He, C.; Zhang, C.; Li, J.; Peng, X.; Meng, L.; Tang, C.; Zhong, J. Direct and Quasi-Direct Band Gap Silicon Allotropes with Remarkable Stability. *Phys. Chem. Chem. Phys.* **2016**, *18* (14), 9682–9686. https://doi.org/10.1039/C6CP00451B.
(32) He, C.; Zhong, J. M585, a Low Energy Superhard Monoclinic Carbon Phase. *Solid State Communications* **2014**, *181*, 24–27. https://doi.org/10.1016/j.ssc.2013.11.035.
(33) Fan, Z.; Chen, W.; Vierimaa, V.; Harju, A. Efficient Molecular Dynamics Simulations with Many-Body Potentials on Graphics Processing Units. *Computer Physics Communications* **2017**, *218*, 10–16. https://doi.org/10.1016/j.cpc.2017.05.003.
(34) Kim, D. Y.; Stefanoski, S.; Kurakevych, O. O.; Strobel, T. A. Synthesis of an Open-Framework Allotrope of Silicon. *Nature Mater* **2015**, *14* (2), 169–173. https://doi.org/10.1038/nmat4140.
(35) Botti, S.; Flores-Livas, J. A.; Amsler, M.; Goedecker, S.; Marques, M. A. L. Low-Energy Silicon Allotropes with Strong Absorption in the Visible for Photovoltaic Applications. *Phys. Rev. B* **2012**, *86* (12), 121204. https://doi.org/10.1103/PhysRevB.86.121204.
(36) Fan, Q.; Chai, C.; Wei, Q.; Yang, Y. Two Novel Silicon Phases with Direct Band Gaps. *Phys. Chem. Chem. Phys.* **2016**, *18* (18), 12905–12913. https://doi.org/10.1039/C6CP00195E.
(37) Zhang, P.; Ouyang, T.; Tang, C.; He, C.; Li, J.; Zhang, C.; Hu, M.; Zhong, J. Thermoelectric Properties of Four Typical Silicon Allotropes. *Modelling Simul. Mater. Sci. Eng.* **2018**, *26* (8), 085006. https://doi.org/10.1088/1361-651X/aae401.
(38) Wei, Q.; Tong, W.; Wei, B.; Zhang, M.; Peng, X. Six New Silicon Phases with Direct Band Gaps. *Phys. Chem. Chem. Phys.* **2019**, *21* (36), 19963–19968. https://doi.org/10.1039/C9CP03128F.
(39) Philipp, H. R.; Taft, E. A. Optical Properties of Diamond in the Vacuum Ultraviolet. *Phys. Rev.* **1962**, *127* (1), 159–161. https://doi.org/10.1103/PhysRev.127.159.
(40) Cai, X. H.; Yang, Q.; Pang, Y.; Wang, M. A Theoretical Prediction of a New Silicon Allotrope: tP36-Si. *Comput. Mater. Sci.* **2020**, *173*, 109441. https://doi.org/10.1016/j.commatsci.2019.109441.
(41) Yu, L.; Zunger, A. Identification of Potential Photovoltaic Absorbers Based on First-Principles Spectroscopic Screening of Materials. *Phys. Rev. Lett.* **2012**, *108* (6), 068701. https://doi.org/10.1103/PhysRevLett.108.068701.

**Table 1.** Numbers of atoms in unit cell (N), space groups (SG), coordination numbers (CN), bandgaps (Eg), energies,transition dipole moments (TDM), Spectroscopic Limited Maximum Efficiencies (SLME) and densities of 22 identified direct bandgap Si structures. The structures that are highlighted in bold were identified in previous studies.

| Si structure | N | SG | CN | Eg (HSE) (eV) | Energy/atom (eV per atom) | TDM2 (Debye2) | Density (g/cm3) | SLME (%) |
|---|---|---|---|---|---|---|---|---|
| 1($Si_{22}$-Pm) | 22 | Pm | 4 | 1.267 | 0.026 | 1.403 | 2.243 | 22.23 |
| **2**[37] | 18 | $P2_1/m$ | 4 | 1.419 | 0.029 | 5.022 | 2.234 | 25.55 |
| 3 | 24 | Pm | 4 | 1.304 | 0.030 | 0.910 | 2.277 | 22.71 |
| 4 | 14 | Pm | 4 | 1.306 | 0.041 | 7.749 | 2.217 | 23.12 |
| 5 | 24 | Pmmn | 4 | 0.882 | 0.046 | 9.981 | 2.274 | 5.99 |
| **6**[38] | 14 | $P2_1/m$ | 4 | 1.183 | 0.048 | 4.866 | 2.219 | 18.19 |
| 7 | 18 | Cm | 4 | 1.384 | 0.056 | 1824.389 | 2.220 | 24.55 |
| 8($Si_{18}$-C2/m) | 18 | C2/m | 4 | 0.796 | 0.065 | 7442.967 | 2.261 | 4.98 |
| **9**[36] | 10 | $P2_1/m$ | 4 | 0.731 | 0.080 | 600.663 | 2.244 | 5.07 |
| 10 | 14 | Pmm2 | 4 | 0.703 | 0.084 | 821.726 | 2.245 | 5.34 |
| 11 | 20 | Cm | 4 | 0.827 | 0.085 | 253.840 | 2.241 | 5.47 |
| 12 | 12 | $P\bar{1}$ | 4 | 1.026 | 0.090 | 87.974 | 2.307 | 18.29 |
| 13 | 12 | Pmmn | 4 | 1.072 | 0.098 | 3.774 | 2.132 | 12.91 |
| 14($Si_{16}$-P2-$_1$/c) | 16 | $P2_1/c$ | 4 | 0.907 | 0.111 | 683.383 | 2.316 | 6.05 |
| 15($Si_{12}$-P1) | 12 | P1 | 4 | 1.694 | 0.116 | 98.219 | 2.110 | 32.28 |
| 16 | 14 | $P\bar{1}$ | 4 | 1.220 | 0.116 | 37.173 | 2.272 | 19.12 |
| **17**[8] | 12 | C2/m | 4 | 1.151 | 0.118 | 38.365 | 2.229 | 17.24 |
| 18 | 18 | C2/m | 4 | 0.949 | 0.131 | 23.285 | 2.087 | 8.57 |
| 19 | 12 | $P\bar{1}$ | 4 | 0.911 | 0.146 | 129.600 | 2.183 | 7.55 |
| 20 | 16 | $P\bar{1}$ | 4 | 1.149 | 0.149 | 26.880 | 2.237 | 16.03 |
| 21 | 12 | C2/m | 4 | 1.404 | 0.176 | 46.113 | 2.098 | 0.00 |
| 22 | 14 | P1 | 4 | 1.411 | 0.223 | 134.482 | 2.010 | 28.29 |
| **Diamond Si** | **2** | **$Fd\bar{3}m$** | **4** | **0.894** | **0.000** | / | **2.313** | **11.20** |

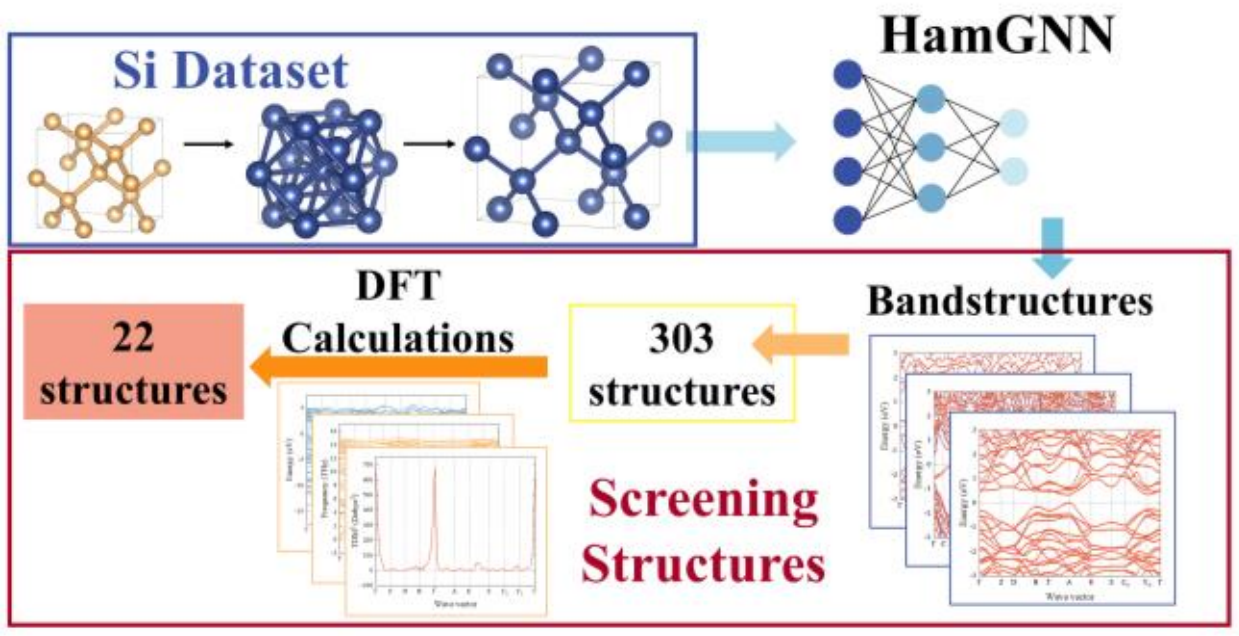

TOC Graphic: Identifying direct bandgap silicon structures with high-throughput search and machine learning methods